\documentclass[fleqn,usenatbib]{mnras}

\usepackage{newtxtext,newtxmath}

\usepackage[T1]{fontenc}

\DeclareRobustCommand{\VAN}[3]{#2}
\let\VANthebibliography\thebibliography
\def\thebibliography{\DeclareRobustCommand{\VAN}[3]{##3}\VANthebibliography}

\usepackage{graphicx}	
\usepackage{amsmath}	

\title[Radio Pulse Profile of Swift J1818.0$-$1607]{Radio Pulse Profile Evolution of Magnetar Swift J1818.0$-$1607}

\author[R. Fisher et al.]{
R. Fisher,$^{1}$\thanks{E-mail: rebecca.fisher-7@postgrad.manchester.ac.uk (RF)}
E. M. Butterworth,$^{1}$ 
K. M. Rajwade,$^{2}$ 
B. W. Stappers,$^{1}$ 
G. Desvignes,$^{3}$ 
R. Karuppusamy,$^{3}$ 
\newauthor
M. Kramer,$^{3}$ 
K. Liu,$^{3}$ 
A. G. Lyne,$^{1}$ 
M. B. Mickaliger,$^{1}$
B. Shaw,$^{1}$
and P. Weltevrede$^{1}$ 
\\
$^{1}$Jodrell Bank Centre for Astrophysics, University of Manchester, Oxford Road, Manchester M13 9PL, UK\\
$^{2}$ASTRON, the Netherlands Institute for Radio Astronomy, Oude Hoogeveensedijk 4, NL-7991 PD Dwingeloo, the Netherlands\\
$^{3}$Max-Planck-Institut für Radioastronomie, Auf dem Hügel 69, D-53121 Bonn, Germany
}

\date{Accepted XXX. Received YYY; in original form ZZZ}

\pubyear{2015}

\begin{document}
\label{firstpage}
\pagerange{\pageref{firstpage}--\pageref{lastpage}}
\maketitle

\begin{abstract}
The shape and polarisation properties of the radio pulse profiles of radio-loud magnetars provide a unique opportunity to investigate their magnetospheric properties.  Gaussian Process Regression analysis was used to investigate the variation in the total intensity shape of the radio pulse profiles of the magnetar Swift J1818.0$-$1607.  The observed profile shape was found to evolve through three modes between MJDs 59104 and 59365.  The times at which these transitions occurred coincided with changes in the amplitude of modulations in the spin-down rate.  The amount of linear and circular polarisation was also found to vary significantly with time.  Lomb-Scargle periodogram analysis of the spin-down rate revealed three possibly harmonically related frequencies.  This could point to the magnetar experiencing seismic activity.  However, no profile features exhibited significant periodicity, suggesting no simple correlations between the profile variability and fluctuations of the spin-down on shorter timescales within the modes. Overall, this implies the mode changes seen are a result of local magnetospheric changes, with other theories, such as precession, less able to explain these observations.
\end{abstract}

\begin{keywords}
stars: neutron -- stars: magnetars -- pulsars: individual: PSR J1818$-$1607.
\end{keywords}



\section{Introduction}
Magnetars are a subclass of neutron stars that are characterised by large inferred magnetic fields (10$^{14}$-10$^{15}$ G), rotation periods of 1-12~s, and characteristic ages less than 200,000 years \citep{KaspiVictoriaM2017M, PulsarAstronomy}.  Magnetars are observed to produce high-energy emission in the form of bursts and persistent X-ray and $\gamma$-ray emission that is thought to be powered by their magnetic fields. This is in contrast to radio pulsars, which are powered by their rotational energy.  Six magnetars have been seen to emit pulses of radio emission, which are much more variable in time than those of standard radio pulsars in terms of flux density, profile shape, and polarisation \citep[e.g.][]{Kramer_2007, Camilo_2007Mag1E, LevinLina2010ARMi, KeithM.J.20111a2o, Lee_2013, ShannonR.M.2013Rpot, LowerMarcusE2020SPoS, Kirsten2021}.   These emission characteristics are also shared with high magnetic field strength pulsars that occasionally display magnetar-like outbursts (e.g., PSR J1119$-$6127, \cite{DAI2018HighBPulsar}).  The onset of radio emission in these objects typically occurs after high-energy outbursts \citep{KaspiVictoriaM2017M} and may therefore be associated with the twisting of the magnetosphere.  The radio emission is thought to originate from the "j-bundle" of field lines \citep{BeloborodovAndreiM2009UMoN}.  

The magnetar Swift J1818.0$-$1607 was discovered in March 2020 via its X-ray outburst \citep{2020X-rayburst} and further observations detected pulsed radio emission \citep{2020ATelJ1818Karup, 2020ATel13554....1R} making it only the fifth radio-loud magnetar \citep{LowerMarcusE2020SPoS}.  
The magnetar has a rotation period of 1.36~s \citep{hu2020a} and therefore has the fastest rotation rate of any magnetar discovered to date.  Previous work on the radio pulse profiles of the magnetar \citep{ChampionDavid2020Hoav, LowerMarcusE2020SPoS, LowerME2021Tdmo,  RajwadeKM2022Ltra} has shown a high variability in their shape and polarisation properties, suggesting it has a very dynamic magnetosphere.  The radio emission was found to be between 80-100 percent linearly polarised and to exhibit a mode change by \cite{LowerMarcusE2020SPoS} and \cite{ChampionDavid2020Hoav}.  This is similar to findings for other magnetars.  However, features such as its radio spectrum being unusually steep early on in its outburst before later flattening to values more typical of magnetars \citep{LowerMarcusE2020SPoS, ChampionDavid2020Hoav} suggest this object may provide a crucial link between the high-magnetic-field radio pulsar and magnetar populations \citep{Hu_2020_SwiftPpr}.  The magnetar 1E 1547.0\(-\)5408 has also been shown to display pulse profile variability that differs from that of normal pulsars, as well as a high percentage of linear polarisation, although this was seen to drop significantly at lower frequencies \citep{Camilo_2007Mag1E}.  The radio pulse profiles of magnetars XTE J1810\(-\)197 \citep{Kramer_2007, CamiloF2007PREf, Camilo2007ApJXTE} and PSR J1622\(-\)4950 \citep{LevinL2012Reep} also have very high levels of linear polarisation, with dramatic variations in shape and polarisation fraction. In the latter, the linear polarisation fraction showed signs of decreasing at lower observing frequencies \citep{LevinLina2010ARMi, LevinL2012Reep}.  Similar behaviour has also been observed in 1E 1547.0\(-\)5408 \citep{Camilo2008}.  Investigating the unusual polarisation characteristics of magnetars is therefore important, as it will provide insight into the radio emission mechanism, magnetospheric properties, and magnetic field geometry of these objects.  

\begin{figure*} 
\centering 
\includegraphics[width=2\columnwidth]{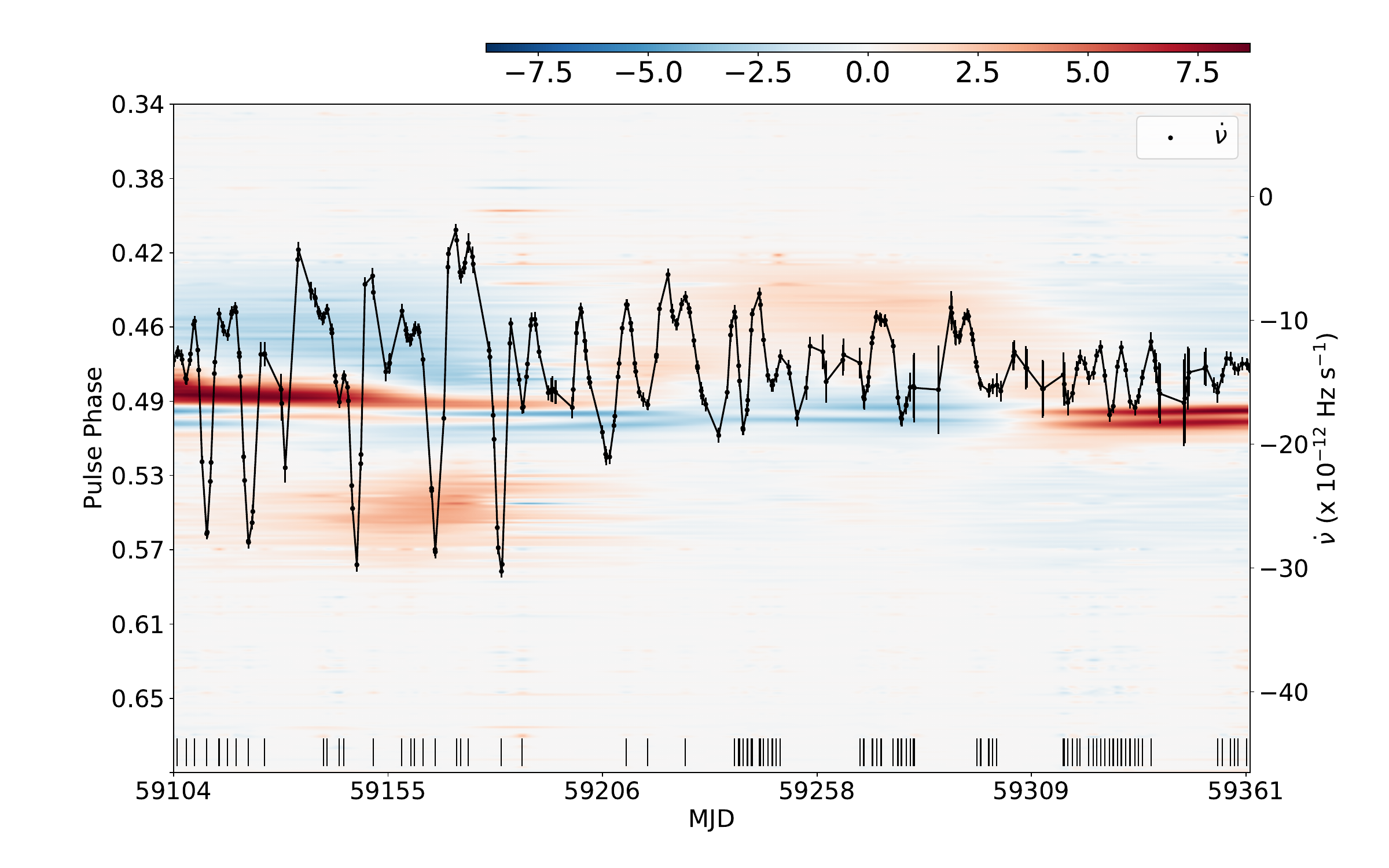}
\caption{Gaussian Process variability map for the total intensity of the pulse profile over time.  The spin-down variation has been overlaid to show that changes in the spin-frequency derivative, \textit{${\dot\nu}$}, modulation amplitude at MJDs 59185 and 59300 coincide with profile shape changes.  The black vertical lines at the bottom of the plot indicate the dates of the observations. 
 The colourbar units are the median of the standard deviation of all the off-pulse regions from the data set, with red indicating an excess in power compared to the median profile and blue a power deficit.} 
\label{GPVariability} 
\end{figure*}

\cite{RajwadeKM2022Ltra} showed that Swift J1818.0\(-\)1607 exhibits an interesting fluctuating spin-down rate (see Fig.~\ref{GPVariability}) and suggested that the magnitude of modulation of the spin-frequency derivative, \textit{${\dot\nu}$} (related to the spin period, $P$, and its time derivative, $\dot P$, as $\dot\nu = -\dot P/P^2$), around the mean value may be correlated with profile shape.  This suggests the magnetic braking torque may be linked to magnetospheric changes.  Regular observations of the radio emission of Swift J1818.0$-$1607 carried out between September 2020 and May 2021 (MJDs 59104-59365) provide the opportunity to study the evolution of the pulse profile characteristics, such as polarisation and shape, and investigate any correlations with the spin-down behaviour, which may provide insight into the underlying physical causes.  The paper is organised as follows: We describe the radio observations of the magnetar in section~\ref{sec:obs}.  Section~\ref{sec:analysis} describes the radio pulse profile calibration method and section~\ref{sec:results} describes the methods and results of the anaylsis used to investigate the evolution of these profiles.  The findings are discussed in section~\ref{sec:discussion} and summarised in section~\ref{sec:summary}.  

\section{Observations}
\label{sec:obs} 
The observations were made using the 76-m Lovell Telescope and the $38\times25$-m Mark II Telescope at the Jodrell Bank Observatory in the UK between MJDs 59104 and 59365 (see Table \ref{tab:example_table} for a full list of the observations used). The central frequency of the observations was 1.53~GHz with a bandwidth of 384~MHz, which was divided into 1532 channels.  Each telescope has two orthogonal dipole receivers in combination with a quarter-wave plate, which converts the signal from linear to circular polarisation.  The voltages are processed using the Digital Filter Backend (DFB) \citep{ManchesterR.N.2013TPPT}, which converts the incoming data into frequency channels and generates Stokes parameters, enabling the linear and circular polarisations to be sampled.  These data are folded onto 1024 phase bins using the best-known ephemeris for the magnetar, and the data were dedispersed using a dispersion measure (DM) of 699~pc~cm$^{-3}$.  These data were then summed into 8-second sub-integrations, resulting in time-frequency-Stokes parameters data cubes that are saved to disk.  The majority of the data analysis was performed using the \textsc{psrchive} \citep{HotanA.W.2004papA} and \textsc{psrsalsa} \citep{WeltevredePSRSALSA} packages.  Channels affected by radio frequency interference (RFI) were removed using \textsc{psrchive}'s {\tt paz}.

\section{Data Analysis}
\label{sec:analysis} 
\subsection{Radio Pulse Profile Polarisation Calibration}
The measured polarisation profiles were calibrated to account for how the telescope's instrumentation responds to the incident radio waves.  Polarisation errors can arise from the gains of the orthogonal dipole receivers being unequal (differential gain), the two feeds not being perfectly orthogonal causing mixing of the orthogonal polarisation components (leakage), imperfections or frequency dependence of the quarter-wave plate, and phase differences due to, for example, different cable lengths for the signals from each feed (differential phase).  Some of these properties of the telescope can vary over the timescale of days.  Calibration observations using a noise diode and well-known sources are used to calibrate pulsar profiles \citep[e.g.][]{vanStratenW2004RAPa, vanStratenW2013HRAP}.  However, at Jodrell Bank, the calibration observations are too infrequent, and so the method of Matrix Template Matching (MTM) \citep{vanStratenW2006RAPa} was used instead. This was performed using the \textsc{psrsalsa} package \citep{WeltevredePSRSALSA}.

\begin{figure} 
\centering 
\includegraphics[width=\columnwidth]{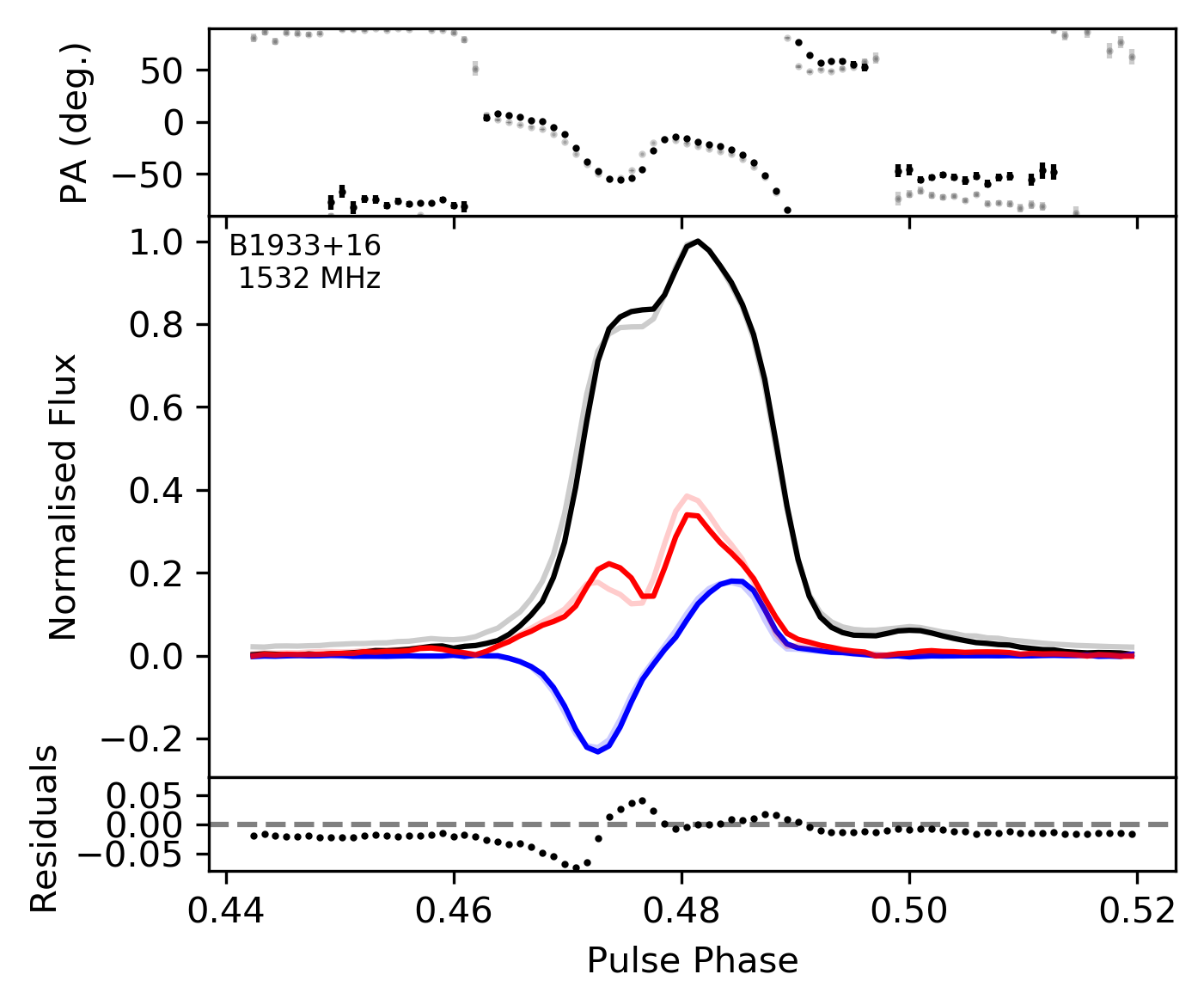} 
\caption{An example of a MTM-calibrated pulse profile from PSR B1933+16 showing the total intensity (black), linear polarisation (red) and circular polarisation (blue) as a function of pulse phase.  The high S/N template \citep{JohnstonSimon2018Po6p} is plotted behind in more transparent colours.  The residuals from subtracting the total intensity of the template from the calibrated profile are shown in the bottom panel.  The calibrated profile shape and PA are sufficiently well matched to the template that any variations in polarisation in the magnetar can be reliably identified.} 
\label{Example_B1933_calib} 
\end{figure}

In MTM, a pulsar with a high percentage of linear polarisation that exhibits stable polarisation properties is observed with the same telescope as the magnetar of interest, and its profile at that time is matched to its high signal-to-noise (S/N) template for a similar frequency, with a separate solution being produced for each frequency channel.  In this case, the bright radio pulsar B1933+16 \citep{MitraDipanjan2016Cacc} was used as it is observed regularly with the Lovell telescope in the same frequency band.  Thus, its frequency-dependent profile could be matched to a high S/N calibrated template \citep{JohnstonSimon2018Po6p} using \textsc{psrsalsa}'s MTM routine {\tt pcal} to obtain calibration solutions over time and frequency.  Linear transformations of the Stokes parameters by Jones matrices are described by Mueller matrices, which have seven independent components \citep{vanStratenW2004RAPa}.  The solutions are determined by fitting a Mueller matrix operating on the Stokes parameters of the observed polarisation and are parameterised in terms of the system's absolute gain, differential phase, and differential gain \citep{HamakerJ.P.1996UrpI}.  The MTM process uses a non-linear least-squares fitting regime to find the parameters that best match the profile to its template.  The resultant Mueller matrix solutions can then be applied to other profiles observed at a similar time and at a similar frequency.

The solutions were reapplied to the dispersion and Faraday rotation corrected B1933+16 pulsar profiles, and the data were dedispersed to produce calibrated profiles with the linear and circular polarisation components, as shown in Fig.~\ref{Example_B1933_calib}.  The profile shape and PA of the calibrated profiles were visually inspected with reference to the high S/N template to check that they were sufficiently well matched to the template such that any variations in polarisation in the magnetar can be reliably identified.

The solutions for the calibrator profile closest in date, with an upper limit of half a day before or after the observation time, were then applied to the Swift J1818.0\(-\)1607 data, initially uncorrected for RM.  The {\tt pcal} routine applies the solutions on a frequency channel-by-channel basis.  Therefore, we added a step in the process to extract the channels with a zero weight for which there was no solution and masked these using \textsc{psrchive}'s {\tt paz} before applying the calibration solutions to ensure only frequency channels available from the calibrator profile were used.  \textsc{psrsalsa}'s {\tt rmsynth} was then used on the calibrated files to obtain the RM for each observation by maximising the degree of linear polarisation \citep{IlieC.D.2019Efme}.

We decided to first use \textsc{psrchive}'s {\tt rmfit} to Faraday rotate the data for RMs of 0-3000 rad m$^{-2}$ in steps of 300 rad m$^{-2}$ and find the value at which the linear polarisation was maximised to get an initial estimate before performing a finer RM search using {\tt rmsynth}.  Using {\tt rmsynth} also provides a more reliable error estimate on the RMs using the method of bootstrapping \citep{WeltevredePatrick2012Pmdb}, whereby white noise with the same root mean square as measured in the off-pulse region is added to each frequency channel and the fitting is performed again for many iterations.  This is more reliable as it makes no assumptions about the signal and can account for non-Gaussian distributed errors \citep{IlieC.D.2019Efme}.  The mean value of the RMs was 1435 $\pm$ 7 rad m$^{-2}$.

The MTM solutions were then applied to the files, and the profiles were corrected using the relevant RM value.  The profiles were dedispersed and a parallactic angle correction was applied.  The shape and polarisation features of the profile on MJD 59109.9 were cross-checked with the profile in \cite{LowerME2021Tdmo} on the same day and were found to be in good agreement.  The total intensity shapes observed are broadly consistent with those shown in \cite{RajwadeKM2022Ltra} and \cite{ChampionDavid2020Hoav}, which both use data from Jodrell Bank but with a different calibration method.

\section{Results}
\label{sec:results} 
\subsection{Radio Pulse Profile Polarisation Evolution}

\begin{figure*} 
\centering 
\includegraphics[width=2\columnwidth]{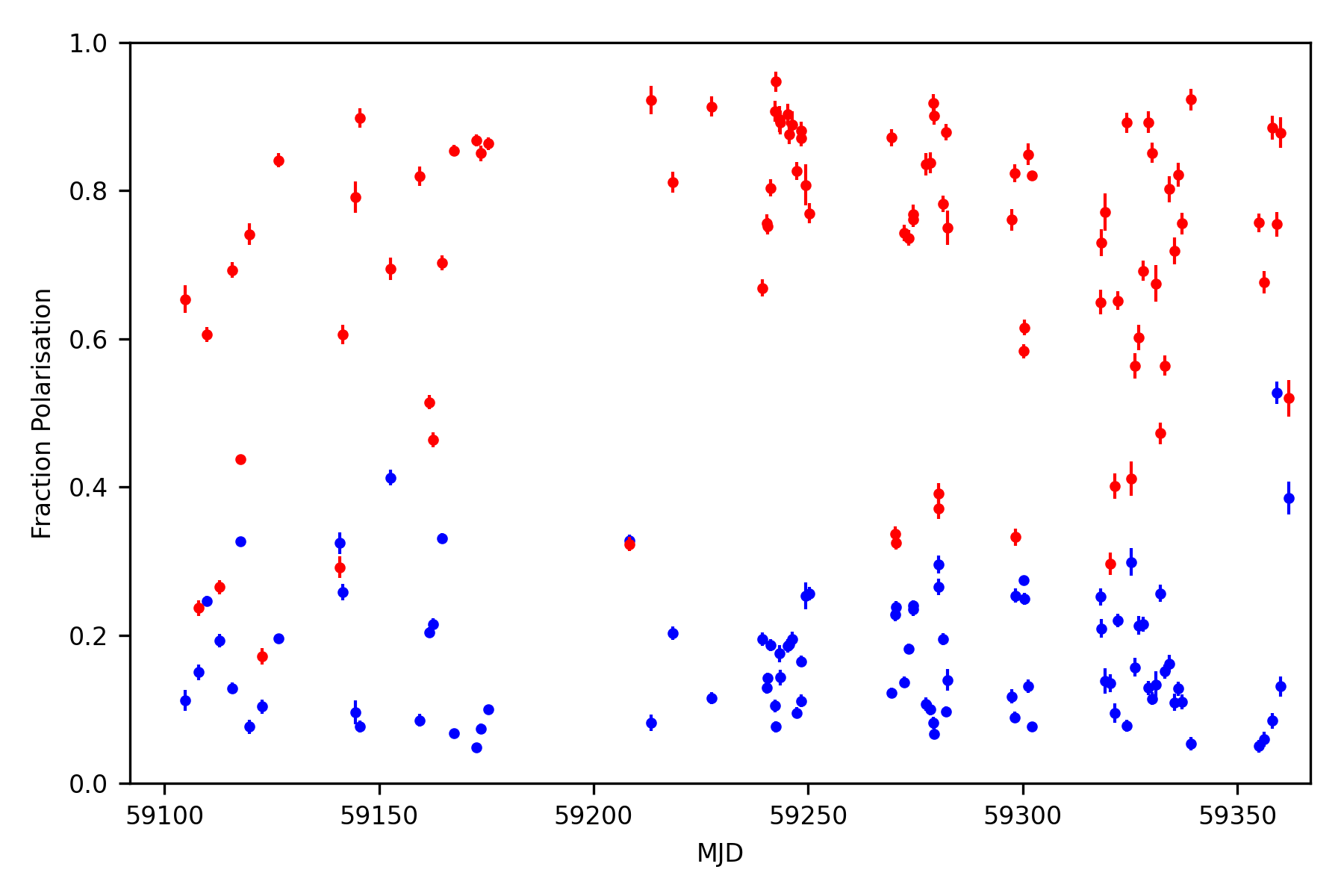} 
\caption{Evolution of the linear (red) and absolute circular (blue) polarisation fractions with MJD.  The linear polarisation was generally high, and most profiles also exhibited significant circular polarisation.} 
\label{Lin_circ_pol_bar} 
\end{figure*}

The linear and circular polarisation fractions were calculated by extracting the profile data and summing the flux density\footnote{Flux densities are not calibrated in this work and are therefore arbitrary.} in the on-pulse bins for each of the total intensity, \textit{I}, linear, \textit{L}, and circular, \textit{|V|}, polarisation flux densities.  The modulus of the circular polarisation was used to avoid cancellation if the handedness changed.  The errors on the fractions were estimated as the root mean square noise in the off-peak bins of the baseline subtracted profile.  The flux density uncertainty, \textit{$\delta F$}, for each phase bin was taken to be the root mean square of the flux in the off-pulse region of the baseline-subtracted profile given by
\begin{eqnarray}
\delta F = \sqrt{\frac{1}{N_{\text{off}} - 1}\sum_{i=1}^{N_{\text{off}}}F_i^{2}},
\end{eqnarray}
where \textit{$F_i$} is the flux density in bin \textit{$i$} and \textit{$N$}\textsubscript{off} is the number of off-pulse bins. Using standard error propagation, the uncertainties on the flux densities in the on-pulse region were calculated.  

The evolution of the linear and circular polarisation fractions is shown in Fig.~\ref{Lin_circ_pol_bar}.  These fractions vary significantly with time, but there was no obvious correlation with the spin-down behaviour of the magnetar.  The main peak of the profile in most cases exhibits a very high fraction of linear polarisation, although there are some exceptions to this.  High linear polarisation fractions in the main peak are common in radio pulsars \citep{GouldD.M.1998Mpo3}, although most pulsars do not show such variability in their polarisation properties over time.  The high linear polarisation fractions are also consistent with both previous observations of this magnetar \citep{ChampionDavid2020Hoav, LowerMarcusE2020SPoS} and the other magnetars with radio emission \citep[e.g.][]{Kramer_2007,CamiloF2007PREf}.  The PA was found to vary considerably in shape, as shown by the example profiles in Fig.~\ref{ModeProfileEx}, and showed significant deviation from the rotating vector model S-like shape seen in many normal pulsars \citep{radhakrishnan1969magnetic}.  When the PA had a constant slope, as in the right-hand panel of Fig.~\ref{ModeProfileEx}, it had a range of gradients, including both positive and negative values.  The PA of the main peak often demonstrated a drop like that shown in the middle panel of Fig.~\ref{ModeProfileEx}, but the magnitude of this drop also varied between observations.  The PA was also seen to have wiggles similar to those seen in the magnetars XTE J1810$-$197 \citep{Kramer_2007, DaiShi2019WPRE} and PSR J1622$-$4950 \citep{LevinL2012Reep} which could hint at propagation effects in the magnetosphere. Further analysis of the PA behaviour will be presented in future work (Liu et al. in prep).   

The majority of the profiles also show significant circular polarisation, but the degree of circular polarisation is almost always lower than the linear polarisation fraction, as seen in many pulsars \citep{RadhakrishnanV.1990Taet}.  The handedness of the circular polarisation is generally the same across each profile, as seen in profiles of other pulsars with high degrees of linear polarisation \citep{HanJ.L.1998Cpip}. 

\subsection{Gaussian Process Regression Analysis}
Gaussian Process Regression (GPR) analysis was used to track the variability of the properties of the calibrated radio pulse profiles over time.  The GPR software {\tt PulsarPVC} \citep{PulsarPVC} aligns the profiles based on their shape using cross-correlation.  However, due to the high variation in the pulse profile shape and uncertainty as to whether features of the pulse profile stayed at the same pulse phase, we decided to modify this to align the profiles using the pulse arrival times to prevent potentially losing this shape information.  This required generating a more precise timing model using the pulsar timing software \textsc{tempo2} \citep{HobbsG.B.2006tanp}, which we used to fit a model to the pulse TOAs based on the magnetar's rotational frequency, \textit{${\nu}$}, and its first six time derivatives.  This gave a phase-connected timing solution for the magnetar over a span of 2 years. In order to do a comparative study of the pulse profile evolution, we decided to align the profiles, which meant whitening the TOAs. To do that, the differences between the actual and predicted TOAs were fitted with 125 sinusoidal waves using the {\tt FITWAVES} method such that the residual of the fit was less than the time span of one pulse phase bin of the profile (spin period (1.36 s) divided by the number of phase bins (1024)) to ensure accuracy to within one bin.  We note that the use of {\tt FITWAVES} is for the sole purpose of roughly aligning the profiles to study the large-scale evolution, and we are insensitive to the daily pulse phase jitter in the profiles once the TOAs are whitened.  The large number of waves needed is due to the magnetar's high levels of timing noise.  We note that this assumes that some fiducial point on the star is being tracked, and is possible because of the high observing cadence of our data that allows us to reliably track the profile components.  The pulse is almost always dominated by a single strong component, which, on short timescales, seems to be fairly stable in phase and has a width that varies between about 2-4\% of the pulse period.  Therefore, we adopted a simple, single-component template of width about 3\% of the pulse period for the entire duration of these observations.  The Mark II profiles were all flagged by {\tt PulsarPVC} as being too low S/N to be included, and so they will not be considered further here.  After alignment, each observed pulse profile was normalised by the mean on-pulse flux density and had the median of all the profiles subtracted from it to generate profile residuals.

A GPR model was generated for the evolution of the profile residuals over time for each phase bin across the pulse profile.  A variability map was then produced, which shows how these differences vary over time with interpolation between observations.  This was performed using {\tt PulsarPVC} \citep{PulsarPVC}, which implements the method described in \cite{Brook2015}.  This uses the Mat\'ern covariance kernel, which can model sharp changes in the profile residuals \citep{RasmussenCarlEdward2005GPfM}, combined with a white noise kernel, which allows the uncertainties, such as the white noise introduced by the telescope's system temperature, in the profile data to be modelled.  The GPR optimises the hyperparameters \textit{$\sigma^2$}, \textit{$\lambda$}, and \textit{$\sigma_n^2$} using a maximum likelihood function.  These hyperparameters describe the signal variance, the covariance lengthscale, and the noise variance of the data, respectively.  The values of these parameters could provide insight into the underlying processes causing profile variability by characterising their temporal behaviour.  We modified the lower limit of the search range for the hyperparameter \textit{$\lambda$} from 30 days to 1 day to allow shorter timescale periodicities to be detected, something that is possible given the high cadence of our data set.

\begin{figure*} 
\centering 
\includegraphics[width=2\columnwidth]{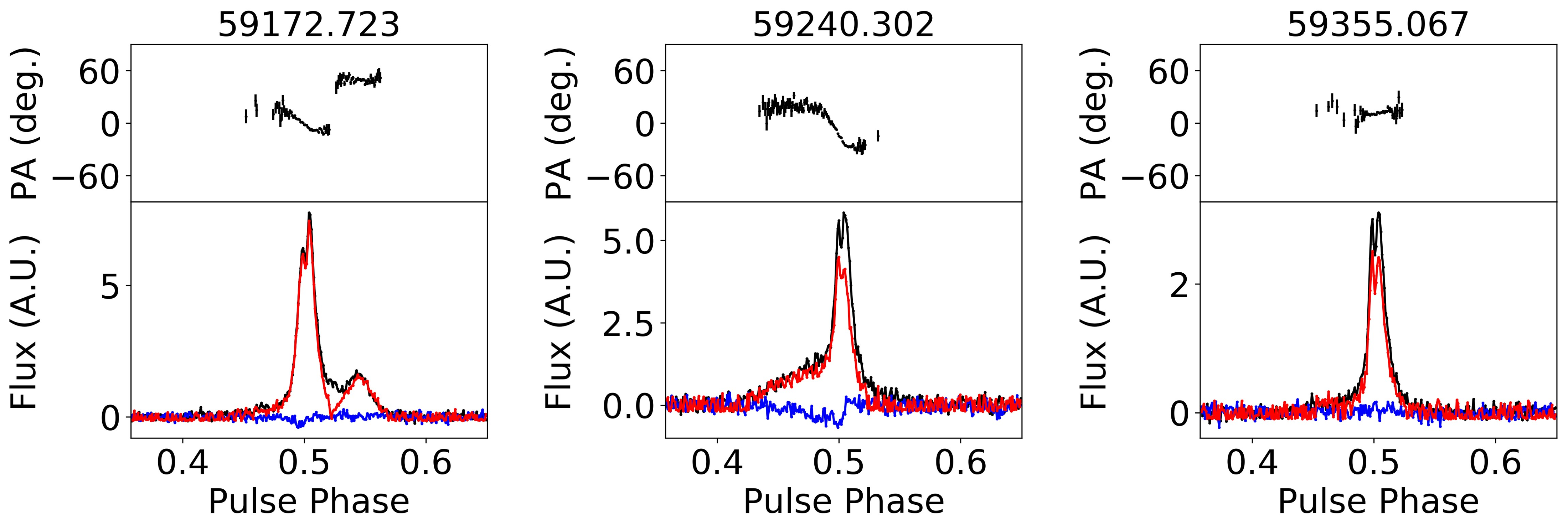}
\caption{Example radio pulse profiles of Swift J1818.0$-$1607 that show the total intensity (black), linear polarisation (red), and circular polarisation (blue) flux density in arbitrary units of the pulse as a function of pulse phase.  The linear polarisation position angles (PAs) are shown in the top panels.  From left to right, the profiles shown are characteristic of the total intensity shapes seen in each of the three modes (MJDs 59104-59184, 59185-59299, 59300-59365) corresponding to the transitions in the amplitude of the spin-frequency derivative, \textit{$\dot\nu$}, modulations, as can be seen in Fig.~\ref{GPVariability}.} 
\label{ModeProfileEx} 
\end{figure*}

The resulting variability map for the pulse profiles' total intensity is shown in Fig.~\ref{GPVariability}.  The colour scale indicates whether the profile has an excess (red) or deficit (blue) of power compared to the median profile.  The colourbar scale is the median of the standard deviation of the off-pulse flux, and the range of values seen indicates the profile shows significant variability.  The median value of \textit{$\lambda$} for pulse phase 0.39-0.59 was 39.5~days, with a median absolute deviation of 9.5~days.  Fig.~\ref{GPVariability} shows that the transitions in modulation amplitude of \textit{$\dot\nu$} at MJDs 59185 and 59300 (seen visually here but quantified by changes in a modulation index defined by \cite{RajwadeKM2022Ltra}) coincide with changes in the shape of the total-intensity radio profile.  An example of a characteristic profile from each of these three modes is shown in Fig.~\ref{ModeProfileEx}.  All plots have the same pulse phase ranges, with the profiles aligned according to the improved timing model.  The profile evolves from having a trailing component that sometimes appears to exhibit an orthogonal-mode-like jump, then a leading component, and finally only exhibiting the main central component.  From Fig.~\ref{GPVariability} it is also evident that there is still variability in the profile shape within the modes, although it is less significant than the variability between modes.

\begin{figure*}
    \centering
    \includegraphics[width=2\columnwidth]{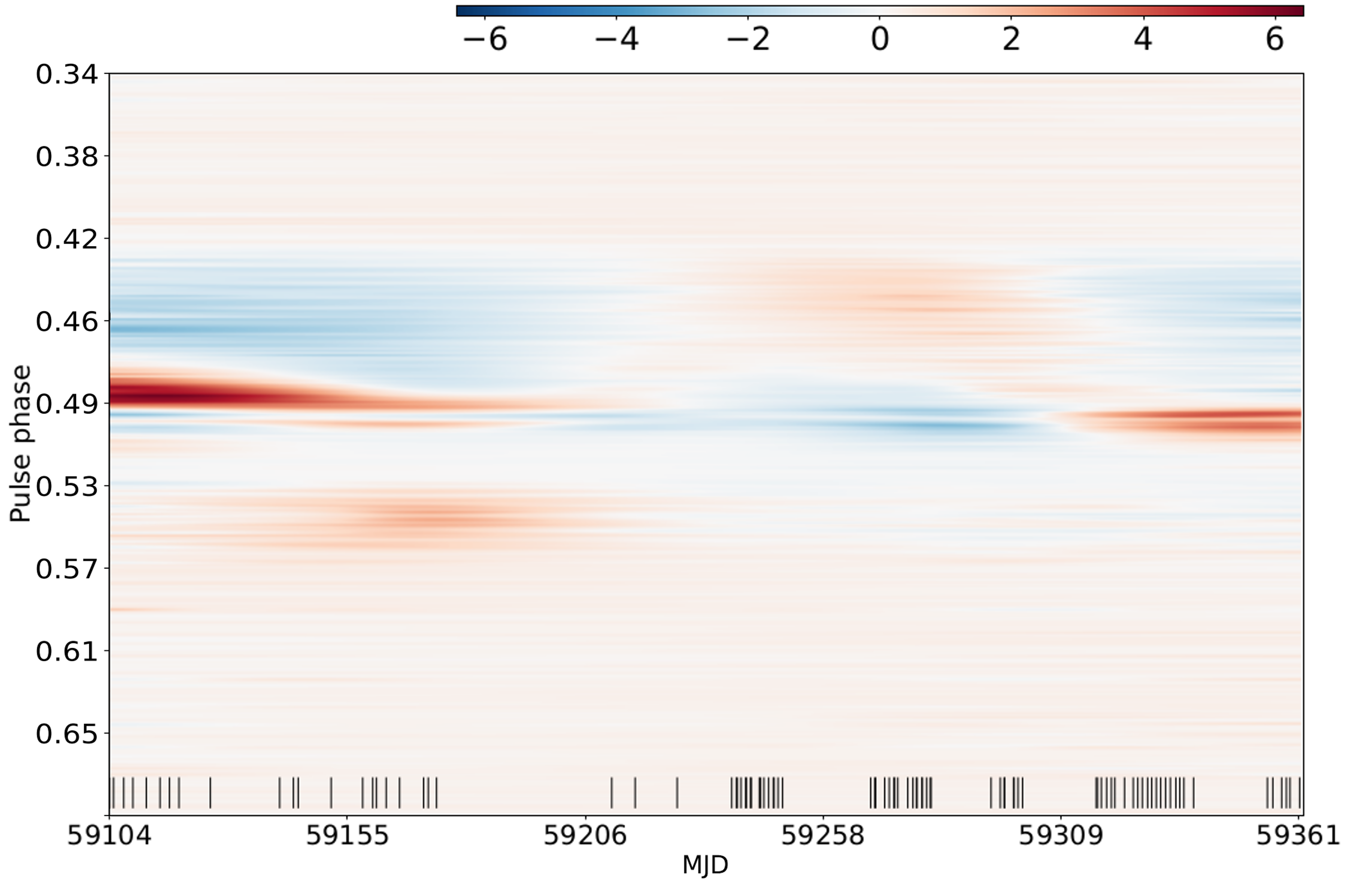}
    \caption{Gaussian Process variability map for the linear polarisation of the pulse profiles. The black vertical lines at the bottom of the plot indicate the dates of the observations.  The colourbar units are the median of the standard deviation of all the off-pulse regions from the data set, with red indicating an excess in power compared to the median linear polarisation profile and blue a power deficit. The similarity between this plot and Fig.\ref{GPVariability} demonstrates the high percentage of linear polarisation generally present in the pulse profiles followed the behaviour of the total intensity changes.}
    \label{LinearPolarisationVariabilityMap}
\end{figure*}

We also modified {\tt PulsarPVC} to run on the linear polarisation component of the profiles.  This produced the variability map shown in Fig.~\ref{LinearPolarisationVariabilityMap}, and it is very similar to that for the total intensity shown in Fig.~\ref{GPVariability}, but with the colours indicating a lower significance (ranging up to a maximum excess of approximately 6 standard deviations) and the transitions being less clear.  This implies that the total and linear intensity components are closely related.  The S/N of the circular polarisation component of the profiles was too low to produce a variability map.   

\subsection{Lomb-Scargle Periodogram Analysis}
\begin{figure*} 
\centering 
\includegraphics[width=2\columnwidth]{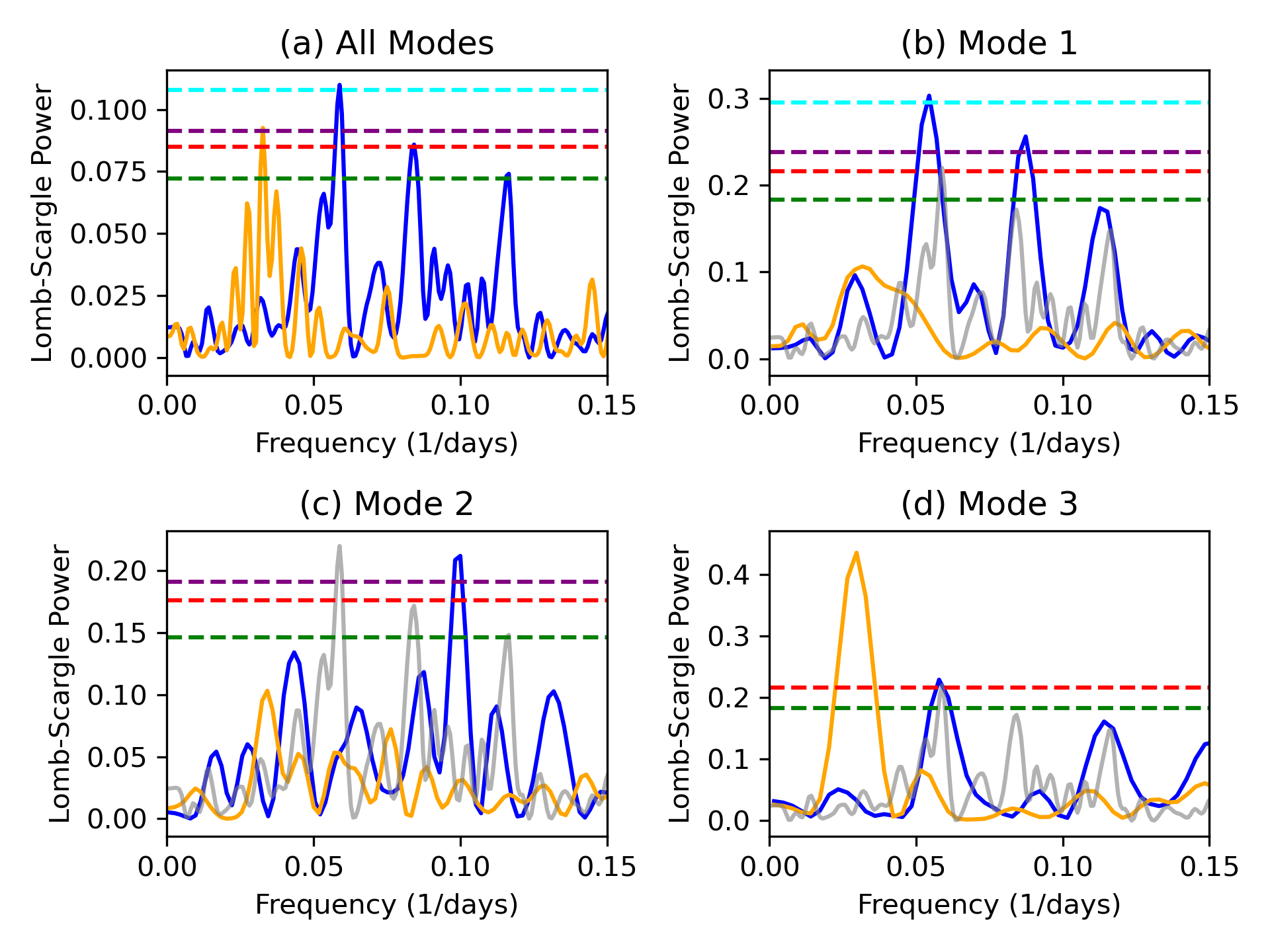}
\caption{Lomb-Scargle periodograms for \textit{$\dot\nu$} are shown in blue for timescales covering all three modes and separately for each mode with their corresponding window functions in yellow.  The 0.3, 0.1, 0.05, and 0.01 FAP levels are shown by the green, red, purple, and cyan dashed lines, respectively.  The three peaks above the 0.3 FAP level in subplot (a) look to be harmonically related.  The periodogram from subplot (a) scaled by a factor of two is shown in grey on the other subplots to show that all three peaks are present again in mode 1 (MJD 59104-59185), two are present in mode 3 (MJD 59300-59365), but in mode 2 (MJD 59185-59300), a new significant frequency emerges.}  
\label{LSModesPlot} 
\end{figure*}
Fig.~\ref{LSModesPlot}(a) shows the Lomb-Scargle periodogram\footnote{This was implemented using the \textsc{astropy} package for \textsc{python} with its default settings \citep{RobitailleThomasP.2013AAcP, Price-WhelanA.M.2018TAPB}.} for \textit{${\dot\nu}$} in blue for the whole period of profile data available (MJDs 59104-59365).  The observed \textit{${\dot\nu}$} signal is a product of the true underlying signal and a window function that describes the observation times.  The window function was generated by creating an array with entries of unity for each date present and plotting it in yellow to allow peaks caused by aliasing or Nyquist-like limits to be identified.  False alarm probabilities (FAPs) calculate the probability that a signal that does not have a periodic component will produce a peak of a given power.  These are used to quantify peak significance.  The most robust way to calculate FAPs is the bootstrap method because it does not make many assumptions about the distribution of the periodogram and accounts for window effects \citep{VanderPlasJacobT2018UtLP}.

As shown in Fig.~\ref{LSModesPlot}(a), three significant frequencies with peaks above the 0.3 FAP that do not coincide with any window function features were identified.  The frequencies of these peaks were 0.059~$\pm$~0.002~days$^{-1}$, 0.084~$\pm$~0.003~days$^{-1}$, and 0.117~$\pm$~0.002~days$^{-1}$.  These peaks look to be harmonically related based on their almost even frequency spacing and amplitude ratios.  The fundamental frequency, however, appears to be missing.
 
The data were then split into the epochs for the three modes (MJDs 59104-59184, 59185-59299, 59300-59365) corresponding to when the amplitude of modulation in \textit{$\dot\nu$} and profile shape changes are seen in the GPR variability plot in Fig.~\ref{GPVariability}.  Lomb-Scargle analysis on \textit{${\dot\nu}$} for each epoch is shown in Fig.~\ref{LSModesPlot}.  All three harmonics are present in the first mode; the second mode develops a new significant frequency at 0.100~$\pm$~0.003~days\textit{$^{-1}$}; and in the third mode, harmonics two and four are present.  This suggests that some of the harmonics persist throughout the data.  

No significant periodicities or periodicities coinciding with the frequencies of the harmonics in \textit{${\dot\nu}$} were found in other properties, such as the linear and circular polarisation fractions.  This suggests that the profile variation on the shorter timescales within the modes is not directly correlated with the spin-down behaviour. 

\section{Discussion}
\label{sec:discussion} 
\subsection{Radio Pulse Profile Polarisation Evolution}
As shown in Fig. \ref{Lin_circ_pol_bar}, the profiles generally have a high degree of linear polarisation and have a median value  of $\sim$17\% circular polarisation across the profiles, both of which vary significantly over time, as is seen in other radio-loud magnetars \citep[e.g.][]{Camilo_2007Mag1E, LevinLina2010ARMi, ShannonR.M.2013Rpot, DaiShi2019WPRE}. This suggests the magnetosphere of the magnetar is very dynamic over a timescale of days.  The main peak in most cases exhibits a very high amount of linear polarisation, although there are some exceptions to this.  The trailing component that develops in the first mode almost always has a significant amount of linear polarisation. The PA associated with this part of the profile gradually develops a jump of about 60 degrees, which was also seen by \cite{ChampionDavid2020Hoav}.  This is an interesting feature, as, for an orthogonal polarisation mode, a jump of 90 degrees would be expected.  Smaller jumps have been observed in other pulsars, and it has been suggested this could be due to the polarisation modes not being completely orthogonal \citep[e.g.][]{SmitsJ.M.2006Fdoo}.  The profiles become depolarised between the two humps, which is consistent with the mixing of modes \citep{PulsarAstronomy}.  A detailed analysis of the PA behaviour will be presented in future work (Liu et al. in prep).  The circular polarisation varies across the pulse phase and could be due to either something intrinsic to the mechanism of radio emission or due to effects during propagation through the magnetosphere \citep{Petrova2000}.  Partially coherent mixing between modes could produce both the observed wiggles seen in the PA and a significant amount of circular polarisation \citep[e.g.][]{Edwards_2004, Oswald2023}.  A correlation between the variation in the linear and circular polarisations could be evidence of Faraday conversion \citep{Edwards_2004, Lower2023} due to propagation in the birefringent magnetosphere plasma surrounding the magnetar.  Excluding profiles from the Mark II telescope and one outlier point with particularly high circular polarisation, the Pearson correlation coefficient between the circular polarisation fraction and one minus the linear polarisation fraction is 0.46 with p-value 3.4$\times10^{-6}$.  This suggests there is not a significant correlation, despite the fact that in some profiles, significant circular polarisation was observed to coincide with a reduction in linear polarisation compared to surrounding profiles.  Further investigation of this possibility is left for future work.  An investigation of this variability demonstrated no correlations between the changes in the polarisation fractions or PA features and other properties, such as the \textit{${\dot\nu}$} variation, of the magnetar.  There also appeared to be no significant periodicity to the variations in these fractions, and the variation could thus not be linked to the mechanism causing the spin-down variation. 

\subsection{Gaussian Process Regression Analysis}
The variability plot produced by the GPR analysis in Fig.~\ref{GPVariability} for the evolution of the pulse profile shows the significant variability of the profile shape.  The high variability of the radio pulse profile shape and the levels of polarisation seen in individual profiles are consistent with those seen in the four other radio-loud magnetars: 1E 1547.0$-$5408 \citep{Camilo_2007Mag1E}, PSR J1622$-$4950 \citep{LevinLina2010ARMi}, SGR J1745$-$2900 \citep{Lee_2013, ShannonR.M.2013Rpot} and XTE J1810$-$197 \citep[e.g.][]{DaiShi2019WPRE}.  The degree of variability over the timescale of days demonstrates the value of high-cadence data in observing magnetars.  If observations were taken less frequently, the level of structure on these short timescales may be undetectable, and, similarly, this also implies that shorter-term variability and patterns may also be missed.

The evolution through the three modes corresponding to the characteristic profile shapes shown in Fig.~\ref{ModeProfileEx} can be seen clearly in Fig.~\ref{GPVariability}.  The overlaid spin-frequency derivative data show that the major changes in profile shape coincide with changes in the amplitude of modulation of the \textit{${\dot\nu}$} variation at MJDs of approximately 59185 and 59300, providing more conclusive evidence for this suggestion than when it was first postulated by \cite{RajwadeKM2022Ltra}.  The variability maps in Fig.~\ref{GPVariability} and Fig.~\ref{LinearPolarisationVariabilityMap} show that the transitions between modes occur suddenly compared to the duration of the modes.    

The mode changes seen in this magnetar are comparable to the phenomenon of mode changing seen in some pulsars.  These pulsars can similarly display periods of quasi-stable emission between sharp discontinuous changes \citep[e.g.][]{Backer1970PeculiarPB, WangN2007}. These changes are very rapid, usually occurring over a single period, and can affect the shape and polarisation of the pulse profile. This dramatic variation is potentially attributable to large-scale redistribution of currents within a pulsar’s magnetosphere, inducing a change to the radio-emission beam pattern and hence a change to the pulse profile \citep[e.g.][]{WangN2007}.  As discussed by \cite{WangN2007}, this suggestion accounts for both the different mode lengths and the short transition timescale: if one of many random perturbations happens to induce positive feedback whereby a small change in magnetospheric currents leads to a change in the electric field, which further enhances this current, rapid large-scale redistribution of currents can result.  \cite{Kramer_2007} found evidence for how changes in plasma density within a magnetosphere at the point of radio emission production may affect the spin-down rate and could lead to changes in the pulse profile.  Emission mode switching has been observed previously in Swift J1818$-$1607 by \cite{LowerME2021Tdmo} and \cite{ChampionDavid2020Hoav}.  These authors note that discrete switching between profile shape modes has been seen to modify the pulse profiles of other magnetars, including affecting the leading edge and causing the development of a secondary trailing component.  Mode switching has also been found to cause a trailing component in the profile of the high-B pulsar PSR J1119$-$6127 \citep{Weltevrede_2011}, which bears resemblance to the changes in the profiles observed here.

The correlation between the modulation amplitude of \(\Dot{\nu}\) and the profile shape changes suggests a link between the spin and the radio emission from the magnetar.  This has been observed in normal pulsars by \cite{LyneAndrew2010SMRo}, \cite{Brook2015}, and \cite{ShawB2022Lrae}, who attributed this to changing plasma densities in the polar cap region.  Changes to plasma density in a magnetosphere, as well as currents within it, can change the torque acting on a pulsar, which affects the spin-down rate \citep{KramerM2006PAPG, LiJ2012ONTS}.  Interestingly, unlike the observations by \cite{ShawB2022Lrae}, we see changes to the size of fluctuations of \(\Dot{\nu}\) around a constant value rather than changes to the absolute value of \(\Dot{\nu}\).  This is perhaps a hint that there are changes to the magnetosphere on different scales, with smaller changes affecting the spin-down but not having a significant effect on the pulse profile.  This may be the case if there are current changes in the magnetosphere close to the light cylinder, which can have a large effect on the torque acting on a magnetar without global magnetospheric changes, which would affect the radio emission \citep{StairsIH2019}.  The changes to the magnetospheric plasma could also be linked to a strong twist that has developed in the magnetosphere and an associated stronger spin-down torque caused by a stronger magnetic field at the light cylinder \citep{Thompson_2002}. The field could subsequently untwist due to crustal movement.  This would result in a decreasing torque and hence reduced amplitude of \(\Dot{\nu}\) and its fluctuations, which are reminiscent of damped oscillations, as seen in this magnetar and others such as XTE J1810\(-\)197 \citep{CamiloF2016RDOT, LevinL2019Sfea}.  The timescales of \textit{${\dot\nu}$} oscillations for normal pulsars are longer than those seen here, which is consistent with magnetars having more dynamic magnetospheres, stronger and more variable currents, and higher rates of pair creation.

Although the correlation between the spin-down and pulse shape changes points to a magnetospheric cause, an alternative explanation is precession.  This would cause the angle between the magnetic dipole and rotation axes of the star to vary periodically and cause the spin-down rate to vary. The pulse profile would also change as a result of a changing line of sight across the beam. In normal pulsars, the expected period of precession is much longer than the changes seen here, and it is expected to be strongly damped by the pinning of superfluid vortices inside the star \citep{LinkB.2006Ioln}, so the magnetospheric explanation is usually preferred \citep{LyneAndrew2010SMRo}. However, it has been suggested that the strong magnetic fields of magnetars could not only deform the star and excite large-amplitude precession but also shatter the crust or suppress the neutron superfluidity, allowing precession to occur \citep{WassermanI2022NPoM}.  Since we do not see profile changes corresponding to each oscillation of \(\Dot{\nu}\), precession seems less able to account for our observations here, and so, a magnetospheric cause seems more likely, but still does not fully explain the lack of profile changes associated with the oscillations.  

\subsection{Lomb-Scargle Periodogram Analysis}
The existence of discrete frequencies in \textit{$\dot\nu$} on timescales shorter than the major mode changes shown in Fig.~\ref{LSModesPlot}(a) that appear to be harmonically related could point to free oscillations caused by seismic activity.  It has been proposed that the outbursts of magnetars could excite toroidal oscillations of the star, which propagate into the magnetosphere, increasing the voltage in the polar cap region and making radio emission more likely \citep{MorozovaViktoriyaS.2012Ereo}.  It is also expected that the oscillations will increase the size of the polar cap and radio emission beam \citep{LinMeng-Xiang2015OMAI}.  The damping of the oscillations means the size of the effect will decay with time.  The decreasing profile width shown in Fig.~\ref{ModeProfileEx} would be consistent with a shrinking polar cap as the toroidal oscillations decay.

The subplots shown in Fig.\ref{LSModesPlot}(b) and (d) show that harmonics two and four persist, perhaps suggesting a common cause or mechanism that persists across the modes.  Fig.\ref{LSModesPlot}(c) shows the emergence of a new significant periodicity.  This could be explained by a change in the dominance of the seismic oscillation modes, which causes magnetospheric changes and thus coincides with the mode changes in the profile shape.  This could be consistent with the suggestion that when the oscillations of seismic modes exceed a critical value, they could cause mode changes in pulsars \citep{LinMeng-Xiang2015OMAI}.  The lack of the fundamental mode frequency is puzzling, as in stellar oscillations, the lower overtone modes are expected to be the most strongly excited.  At present, there are no theoretical predictions as to whether seismic oscillations with frequencies of days could be produced based on predictions of the structure of magnetars.

Although we conclude above that precession is an unlikely explanation for our observations, we note that it could provide an alternative explanation for the persisting periodicity since it would cause strong coupling between magnetospheric variations and the spin-down torque.  The precession period is of order \textit{$P/\epsilon$}, where \textit{$P$} is the magnetar's spin period and \textit{$\epsilon$} is the star's ellipticity \citep{WassermanI2022NPoM}.  Upper limits for \textit{$\epsilon$} considering the elasticity of the crust and magnetic deformation \citep{J1810Precession} result in a prediction for the precession period of around 16~days and 252~days, respectively.  The elastic crust strain \textit{$\epsilon$} could give a precession period on the correct timescale for the periodicities seen.  Precession has been invoked to explain 16-day periodicities in a fast radio burst, which are thought to be related to magnetars \citep{Tong_2020}.

Quasi-periodic variations in \textit{$\dot\nu$} are seen in other magnetars such as 1E 1048.0$-$5937 \citep{ArchibaldR.F.2015Rdtv} and also in some rotation-powered radio pulsars \citep{KramerM2006PAPG, LyneAndrew2010SMRo, ShawB2022Lrae} on longer timescales.  The spin-down torque in untwisting magnetar magnetospheres is expected to vary in a non-monotonic way \citep{BeloborodovAndreiM2009UMoN}, which could account for the fluctuating behaviour of \textit{$\dot\nu$}.  However, it has been argued by \cite{TongH2013WBOM} that short-term spin-down variations in magnetar PSR J1622$-$4950, on similar timescales to those seen here in Swift J1818.0$-$1607, are hard to explain using the theory of magnetospheric untwisting \citep{BeloborodovAndreiM2009UMoN} and could instead be explained by a particle wind.  In the particle wind braking model, it is predicted that the spin-down would exhibit behaviour in the higher-order derivatives \citep{TongH2013WBOM} due to the torque from the distortion of the dipole field near the light cylinder affecting the spin-down of the magnetar \citep{HardingAliceK1999MS}.  This explanation could also link to seismic aftershocks, which can produce variations in the particle wind \citep{ThompsonChristopher1996TSGR}, allowing it to vary significantly on shorter timescales and cause timing noise.  This mechanism has been suggested to explain the timing noise of normal pulsars, in which case the wind is rotation-powered \citep{LyneAndrew2010SMRo}, but it could be powered by the decay of the magnetic field of the magnetar \citep{TongH2013WBOM}.  This mechanism of wind braking is thought to not change the global dipole field of the magnetar \citep{TongH2013WBOM}, and this could explain why the overall profile shape stays reasonably stable within each mode.  The modes themselves could instead be a result of a global magnetospheric reconfiguration \citep{ShawB2022Lrae}.  Further supporting evidence for the wind braking model would be the presence of a pulsar wind nebula powered by the magnetar.  Extended diffuse emission has been detected in X-ray and radio for Swift J1818.0$-$1607 \citep{IbrahimA.Y.2023DXaR}, but there is no strong evidence for this being a nebula, and it may be a dust scattering halo. 

While the magnitude of the oscillations in \textit{$\dot\nu$} is clearly correlated with the pulse profile mode changes, the lack of coinciding periodicities in any of the Lomb-Scargle periodogram analyses on all other profile parameters suggests variability within the profile modes is not strongly coupled to the \textit{$\dot\nu$} oscillations or the underlying mechanism.  In the study of the pulse profile variability of normal pulsars in \cite{ShawB2022Lrae}, it was suggested that more subtle changes in the profile caused by smaller magnetospheric changes would not be able to affect \textit{$\dot\nu$} in an observable way.  While the pulse variability within each profile mode did not exhibit periodic behaviour, the variability plot in Fig.\ref{GPVariability} does seem to show there was a greater degree of variability earlier on.  For example, in mode 1, there is the appearance of the trailing profile component, and there is variability in the leading slope in mode 2, compared to a much more consistent narrow peak in mode 3.  This could imply that the greater the width or complexity of the profile, and hence the radio-emitting region, the greater the magnitude and complexity of the effects on the spin-down torque.  Complex interactions, delays, or feedback could mean that there is no simple relationship between profile characteristics, and spin-down variations may be the cumulative effect of many processes.  

\section{Summary}
\label{sec:summary} 
The radio emission properties of the magnetar Swift J1818.0$-$1607 seem to be in agreement with many of the suggested explanations for magnetar emission.  However, it also demonstrates complex, highly variable behaviour that is not fully understood.  The onset of radio emission after an X-ray outburst and the characteristics such as the decaying flux and wide beam are in agreement with emission from the j-bundle in a magnetosphere that is twisted by seismic activity in the neutron star crust and subsequently untwists \citep{BeloborodovAndreiM2009UMoN}.  The clear correlation between the major profile shape mode changes and the amplitude of the spin-frequency derivative variations shown in Fig.\ref{GPVariability} is in strong support of a twisted magnetic field exerting a torque on the magnetar \citep{Thompson_2002}.  The narrowing profile is consistent with a shrinking emission region and suggests these profile shapes reflect emission from different regions of the beam.  The presence of harmonically related frequencies from the Lomb-Scargle analysis of \textit{$\dot\nu$} shown in Fig.\ref{LSModesPlot} could potentially point to seismic oscillation modes \citep{MorozovaViktoriyaS.2012Ereo, LinMeng-Xiang2015OMAI} and may also link to a particle wind that could cause both timing noise and the emission to vary \citep{TongH2013WBOM}.  The high variability in the pulse profile shape within the modes and \textit{$\dot\nu$} points to a dynamic magnetosphere.  The lack of correlation between the \textit{$\dot\nu$} variation and profile features within the modes could be explained either by the fact that the variability in the magnetosphere can't produce observable changes \citep{ShawB2022Lrae} or that the processes are too complex to produce simple correlations. 

\section*{Acknowledgements}
Access to the Lovell telescope and the Mark II telescope is supported by a consolidated grant from the UK Science and Technology Facilities Council (STFC).  K.M.R. acknowledges support from the Vici research program `ARGO' with project number 639.043.815, financed by the Dutch Research Council (NWO).  The authors would like to thank the anonymous referee, whose comments significantly improved the manuscript. 
\section*{Data Availability}
The data used in this manuscript will be made available to others upon reasonable request to the authors.


\bibliographystyle{mnras}
\bibliography{example_new} 




\appendix

\section{Observations}

\begin{table*}
	\centering
	\caption{Table of observations used with their telescope, start time, duration, and MJDs.  All observations had a central frequency of 1.53~GHz and a bandwidth of 384~MHz.}
	\label{tab:example_table}
	\begin{tabular}[t]{lccr} 
		\hline
		Telescope & Time Start (hhmmss) & Duration (s) & Topocentric MJDs\\
		\hline
		Lovell &  172854 & 1238.8 &  59105.729 \\
        Lovell &  215627 & 1388.9 &  59107.914 \\
        Lovell &  212904 & 728.5 &  59109.895\\
        Lovell &  190442 & 538.9 &  59112.795\\
        Lovell &  181407 & 888.2 &  59115.76\\
        Lovell &  191633 & 798.1 &  59117.803\\
        Lovell &  203052 & 719.0 &  59119.855\\
        Lovell &  195441 & 709.4 &  59122.83\\
        Lovell &  153315 & 1828.2 &  59126.648\\
        Lovell &  194549 & 708.1 &  59140.824\\
        Lovell &  144259 & 708.1 &  59141.613\\
        Lovell &  140400 & 708.1 &  59144.586\\
        Lovell &  145522 & 698.6 &  59145.622\\
        Lovell &  174555 & 719.1 &  59152.74\\
        Lovell &  132013 & 908.7 &  59159.556\\
        Lovell &  173250 & 719.1 &  59161.731\\
        Lovell &  125129 & 709.5 &  59162.536\\
        Lovell &  151647 & 668.6 &  59164.637\\
        Lovell &  124941 & 1837.9 &  59167.535\\
        Lovell &  172028 & 708.2 &  59172.723\\
        Lovell &  173026 & 719.1 &  59173.73\\
        Lovell &  122059 & 708.2 &  59175.515\\
        Lovell &  100251 & 708.2 &  59208.419\\
        Lovell &  090335 & 719.1 &  59213.378\\
        Lovell &  103252 & 717.7 &  59218.44\\
        Lovell &  123143 & 709.5 &  59227.522\\
        Lovell &  071435 & 698.6 &  59239.302\\
        Lovell &  071401 & 659.0 &  59240.302\\
        Lovell &  115849 & 698.6 &  59240.499\\
        Lovell &  072117 & 728.6 &  59241.307\\
        Lovell &  074802 & 717.7 &  59242.325\\
        Lovell &  111543 & 719.1 &  59242.469\\
        Lovell &  074816 & 349.3 &  59243.325\\
        Lovell &  125634 & 719.1 &  59243.539\\
        Lovell &  062221 & 627.7 &  59245.266\\
        Lovell &  123158 & 708.2 &  59245.522\\
        Lovell &  060648 & 518.5 &  59246.255\\
        Lovell &  065138 & 709.5 &  59247.286\\
        Lovell &  065823 & 709.5 &  59248.291\\
        Lovell &  104636 & 717.7 &  59248.449\\
        Lovell &  074945 & 148.7 &  59249.326\\
        Lovell &  070637 & 566.3 &  59250.296\\
        Mark II &  062034 & 1258.1 &  59252.264\\
        Mark II &  113909 & 2297.8 &  59252.486\\
        Mark II &  061245 & 1318.1 &  59254.259\\
        Mark II &  112329 & 2139.5 &  59256.475\\
        Mark II &  100054 & 2948.7 &  59257.417\\
        Mark II &  112823 & 2258.2 &  59260.478\\
        Mark II &  054508 & 3478.0 &  59261.24\\
        Lovell &  081627 & 608.6 &  59269.345\\
        Lovell &  053719 & 708.2 &  59270.235\\
        Lovell &  094920 & 708.2 &  59270.41\\
        Mark II &  050553 & 1379.5 &  59271.213\\
        Mark II &  102433 & 3498.5 &  59271.434\\
        Mark II &  050156 & 1667.4 &  59272.21\\
        Lovell &  083934 & 708.2 &  59272.361\\
        Mark II &  045802 & 3388.0 &  59273.207\\
        Lovell &  082102 & 709.5 &  59273.348\\
        Mark II &  045406 & 3468.5 &  59274.204\\
        Lovell &  092504 & 717.7 &  59274.393\\
        Lovell &  111803 & 709.5 &  59274.471\\
    \end{tabular}
    \begin{tabular}[t]{lccr} 
		\hline
		Telescope & Time Start (hhmmss) & Duration (s) & Topocentric MJDs\\
		\hline
        Mark II &  045009 & 3468.5 &  59275.202\\
        Mark II &  100843 & 3329.3 &  59275.423\\
        Lovell &  082200 & 399.8 &  59277.349\\
        Lovell &  110739 & 708.2 &  59278.464\\
        Lovell &  050211 & 709.5 &  59279.21\\
        Lovell &  083701 & 719.1 &  59279.359\\
        Lovell &  100059 & 709.5 &  59280.418\\
        Lovell &  110246 & 719.1 &  59280.46\\
        Lovell &  095129 & 708.2 &  59281.411\\
        Lovell &  043035 & 698.6 &  59282.188\\
        Lovell &  093405 & 148.7 &  59282.399\\
        Mark II &  034717 & 3233.9 &  59291.158\\
        Mark II &  090549 & 3498.6 &  59291.379\\
        Mark II &  034319 & 3468.6 &  59292.155\\
        Mark II &  090148 & 3508.1 &  59292.377\\
        Mark II &  043829 & 3197.0 &  59293.194\\
        Mark II &  085746 & 839.2 &  59293.374\\
        Mark II &  091558 & 1717.9 &  59294.386\\
        Mark II &  033136 & 1195.3 &  59295.147\\
        Mark II &  084959 & 3508.2 &  59295.368\\
        Mark II &  040653 & 3609.1 &  59296.172\\
        Mark II &  094554 & 3449.5 &  59296.407\\
        Lovell &  080700 & 296.1 &  59297.338\\
        Lovell &  043322 & 708.2 &  59298.19\\
        Lovell &  072543 & 551.3 &  59298.31\\
        Lovell &  045017 & 719.1 &  59300.202\\
        Lovell &  074959 & 708.2 &  59300.327\\
        Lovell &  042849 & 278.4 &  59301.187\\
        Lovell &  041139 & 3609.2 &  59302.175\\
        Mark II &  024813 & 839.2 &  59306.117\\
        Mark II &  080637 & 3528.7 &  59306.338\\
        Lovell &  014405 & 709.6 &  59318.072\\
        Lovell &  075539 & 708.2 &  59318.331\\
        Lovell &  041947 & 237.4 &  59319.181\\
        Lovell &  074621 & 708.2 &  59320.324\\
        Lovell &  082259 & 698.7 &  59321.35\\
        Lovell &  024033 & 1277.2 &  59322.112\\
        Lovell &  043538 & 578.6 &  59324.192\\
        Lovell &  075601 & 708.2 &  59325.331\\
        Lovell &  015330 & 708.2 &  59326.079\\
        Lovell &  012506 & 436.7 &  59327.059\\
        Lovell &  010308 & 708.2 &  59328.044\\
        Lovell &  042253 & 717.8 &  59329.183\\
        Lovell &  014409 & 717.8 &  59330.073\\
        Lovell &  012657 & 719.1 &  59331.061\\
        Lovell &  001204 & 719.1 &  59332.009\\
        Lovell &  012607 & 698.7 &  59333.06\\
        Lovell &  021847 & 719.1 &  59334.097\\
        Lovell &  065152 & 719.1 &  59335.286\\
        Lovell &  031848 & 687.8 &  59336.138\\
        Lovell &  003926 & 719.1 &  59337.028\\
        Lovell &  024258 & 709.6 &  59339.113\\
        Lovell &  013630 & 708.3 &  59355.067\\
        Lovell &  054600 & 698.7 &  59356.241\\
        Lovell &  034107 & 717.8 &  59358.154\\
        Lovell &  040406 & 719.2 &  59359.17\\
        Lovell &  230258 & 709.6 &  59359.961\\
        Lovell &  005248 & 708.3 &  59362.037\\

		\hline
	\end{tabular}
\end{table*}


\bsp	
\label{lastpage}
\end{document}